\documentclass{IMAGE2025}
\usepackage{amsmath}
\usepackage{hyperref}
\usepackage{array}  
\usepackage{amsmath, amssymb, amsthm}
\newtheorem{definition}{Definition}

\ifdefined\CSLReferences\else
\newenvironment{CSLReferences}[2]{%
  \begin{list}{}%
    {%
      \setlength{\leftmargin}{0em}%
      \setlength{\itemindent}{-\leftmargin}%
      \setlength{\itemsep}{\parsep}%
      \setlength{\parsep}{0pt}%
      \setlength{\topsep}{0pt}%
    }%
  \item[] 
}{%
  \end{list}
}
\fi

\begin{document}

\title{A reduced-order derivative-informed neural operator for subsurface fluid-flow
}

\renewcommand{\thefootnote}{\fnsymbol{footnote}} 

\author{Jeongjin Park, Grant Bruer, Huseyin Tuna Erdinc, Abhinav Prakash Gahlot, and Felix J. Herrmann \\
Georgia Institute of Technology}

\maketitle

\begin{abstract}
Neural operators have emerged as cost-effective surrogates for expensive fluid-flow simulators, particularly in computationally intensive tasks such as permeability inversion from time-lapse seismic data, and uncertainty quantification. In these applications, the fidelity of the surrogate’s gradients with respect to system parameters is crucial, as the accuracy of downstream tasks, such as optimization and Bayesian inference, relies directly on the quality of the derivative information. Recent advances in physics-informed methods have leveraged derivative information to improve surrogate accuracy. However, incorporating explicit Jacobians can become computationally prohibitive, as the complexity typically scales quadratically with the number of input parameters. To address this limitation, we propose DeFINO (Derivative-based Fisher-score Informed Neural Operator), a reduced-order, derivative-informed training framework. DeFINO integrates Fourier neural operators (FNOs) with a novel derivative-based training strategy guided by the Fisher Information Matrix (FIM). By projecting Jacobians onto dominant eigen-directions identified by the FIM, DeFINO captures critical sensitivity information directly informed by observational data, significantly reducing computational expense. We validate DeFINO through synthetic experiments in the context of subsurface multi-phase fluid-flow, demonstrating improvements in gradient accuracy while maintaining robust forward predictions of underlying fluid dynamics. These results highlight DeFINO's potential to offer practical, scalable solutions for inversion problems in complex real-world scenarios, all at substantially reduced computational cost.
\end{abstract}

\subsection{INTRODUCTION}\label{introduction}

Employing neural networks as surrogate models to approximate solutions of parametric PDEs has shown promise in addressing many-query scenarios such as Bayesian inference, optimization under uncertainty, and Bayesian optimal experimental design \cite[]{qiu2024derivative, o2024derivative}. Such problems typically rely on complex mathematical models, often parameterized PDEs, which require numerous forward model evaluations, i.e., solving the PDEs for many parameter settings, making high-fidelity simulation computationally demanding \cite[]{gao2024adaptive, sheriffdeen2019accelerating, wu2024uncertainty}. To overcome this challenge, neural operator surrogates, particularly Fourier Neural Operators (FNOs), have emerged as efficient alternatives for approximating PDE solutions \cite[]{yin2022learned, li2020fourier, wen2022u, chandra2025fourier}. By leveraging computationally inexpensive approximations of the forward model, these surrogates significantly reduce the computational burden, thus enabling more scalable and practical downstream tasks, such as inversion problems, and uncertainty quantification \cite[]{yin2023solving}.

When neural operators are used as surrogates for numerical simulators or governing equations of physical systems, it is essential that they adhere to fundamental physical principles, such as mass conservation, entropy decay, and invariant measures \cite[]{li2024physics, park2024dynamical, cai2021physics}. However, standard neural network training approaches, typically based on mean squared error (MSE) objectives, may struggle to accurately learn true physics, limiting the reliability and generalizability of the resulting surrogate models \cite[]{park2024dynamical, rosofsky2023applications, li2024physics}. For instance, FNOs have exhibited instabilities in long-term predictions, particularly for stiff systems, systems that evolve on vastly different time scales, highlighting the need for robust training strategies capable of faithfully capturing complex physical behaviors \cite[]{park2024dynamical, li2024physics}.

Such inaccuracies can significantly compromise the robustness and reliability of subsequent downstream tasks, such as Bayesian inversion, optimization under uncertainty, and Bayesian optimal experimental design, for which a precise gradient is critical. To address these limitations, particularly for inherently complex problems such as chaotic systems (e.g., weather forecasting) or high-dimensional time-dependent fluid-flow scenarios, recent studies have introduced derivative-based training approaches. Approaches such as Physics-Informed Neural Networks (PINNs), and Derivative-Informed Neural Operators (DINOs) integrate derivative-based penalty terms, PDE residual constraints, and higher-order derivative information directly into the training process \cite[]{li2024physics, park2024dynamical, o2024derivative, wu2023large}. The primary motivation behind these derivative-informed strategies is to enhance surrogate models' ability to accurately represent not only input-output relationships but also underlying sensitivities and dynamical behaviors governed by PDE systems \cite[]{o2022derivative}. Nevertheless, applying these methods to complex, high dimensional real-world problems, such as subsurface fluid-flow, remains challenging due to stringent computational requirements, including fully differentiable simulators and substantial memory and runtime demands \cite[]{o2024derivative}.

\subsubsection{Learning derivatives with reduced-order modeling}

One of the major obstacles in physics-based machine learning, particularly for high-dimensional systems, is the high computational cost required to compute derivatives. This renders many conventional derivative estimation methods impractical for large-scale, real-world problems \cite[]{o2022derivative}. Thus, researchers have employed dimension reduction techniques, such as Principal Component Analysis (PCA) or Active Subspace, to compress the input or output into an efficient lower-dimensional representation, resulting in so-called \textit{reduced-order modeling} \cite[]{o2022derivative, constantine2014active, benner2015survey, quarteroni2015reduced}. However, despite their advantages, reduced-order models, which reduce the dimensions of input-output, may encounter limitations when applied to highly nonlinear or multi-scale systems \cite[]{constantine2015active, benner2015survey, quarteroni2015reduced}. That is, such approaches risk discarding critical information if essential features are not captured within the reduced subspace, which may lead to inaccurate derivative approximations.

Our approach addresses these limitations by computing matrix-free actions of the Jacobian without reducing the dimensionality of the input or output, thus maintaining the full-order structure of the problem. That is, our training algorithm encourages the neural operator to approximate full-order Jacobians in a scalable manner by leveraging eigenvectors derived from the Fisher Information Matrix (FIM), as defined in Definition~\ref{def:FIM}. \\

\begin{definition}[Fisher Information Matrix]
\label{def:FIM}
Let $\mathbf{A}$ denote the input parameters and $\mathbf{Y}$ represent the observations. Given a likelihood function $p(\mathbf{Y} \mid \mathbf{A})$, the Fisher Information Matrix (FIM), denoted by $I(\mathbf{A})$, is defined as:
\begin{equation}
\label{eq:FIM}
I(\mathbf{A}) = \mathbb{E}_\mathbf{Y}\left[ \left(\nabla_\mathbf{A} \log p(\mathbf{Y} \mid \mathbf{A})\right)^\top \left(\nabla_\mathbf{A} \log p(\mathbf{Y} \mid \mathbf{A})\right)\right].
\end{equation}
\end{definition}

Unlike traditional data-driven dimension reduction methods such as Principal Component Analysis (PCA) or Proper Orthogonal Decomposition (POD), the FIM offers a gradient-based approach directly informed by the likelihood function \cite[]{quarteroni2015reduced}. Furthermore, in contrast to other gradient-based methods like Active Subspace, which typically rely on gradient evaluations obtained by sampling the parameter space according to a predefined prior, the FIM inherently incorporates observational uncertainty through the likelihood function itself \cite[]{constantine2014active}, \cite[]{zahm2022certified}, \cite[]{cui2021data}. Consequently, in our context for the subsurface fluid-flow problem, where the underlying distribution of permeability models is unknown \cite[]{gahlot2024uncertainty}, the FIM provides a more suitable and robust basis for dimension reduction.

\subsubsection{Our contribution}

To address these challenges, we propose a Derivative-based Fisher-score Informed Neural Operator (DeFINO), a novel framework designed to enhance neural operators by explicitly incorporating reduced-order Jacobian information into the training process. DeFINO differs from recent derivative-informed modeling approaches \cite[]{o2022derivative, o2024derivative, zahm2022certified, qiu2024derivative} by maintaining the surrogate model in its full-order representation, rather than reducing the input or output dimensionality directly. Instead, DeFINO achieves dimension reduction by projecting the Jacobian onto dominant eigenvectors of the FIM. This strategy focuses the surrogate model's training on the parameter directions most sensitive to observational data, enabling accurate yet computationally efficient derivative approximations, particularly beneficial for high dimensional problems.

Our primary contributions include: (1) introducing a novel derivative-informed training strategy for neural operators applicable to high-dimensional systems; (2) developing a scalable method to compute the FIM, including deriving a closed-form expression for the practical scenario of a multivariate Gaussian likelihood; and (3) validating the effectiveness of the DeFINO framework through numerical experiments. We specifically evaluate DeFINO's performance on a subsurface fluid-flow problem, highlighting its effectiveness even when the available number of training samples is limited.

\subsection{METHODOLOGY}\label{methodology}

\subsubsection{Problem description: empirical risk minimization}

In this work, our goal is to improve the surrogate model's prediction of subsurface fluid-flow behavior while accurately capturing gradients with respect to the input permeability field. The plume dynamics are modeled by two-phase flow equations, which we express as the mapping $\mathbf{A} \rightarrow \mathbf{U} = \mathcal{F}(\mathbf{A})$, where $\mathcal{F}$ represents the two-phase flow simulation that models the time-lapse of CO$_2$ concentration snapshots $\mathbf{U} = [\mathbf{u}_1, \mathbf{u}_2, \cdots \mathbf{u}_k]$ given the permeability field $\mathbf{A}$. Our proposed training strategy is formalized the following empirical risk minimization problem:
\begin{equation}
\begin{aligned}
\min_{\boldsymbol{\theta}} \frac{1}{N}\sum_{j=1}^N \Biggl[ 
&\|\mathcal{F}(\mathbf{A}_j) - \mathcal{F}_{nn}(\mathbf{A}_j; \mathbf{\theta})\|^2_2 \\
&+ \lambda \sum_{i=1}^r \|\mathbf{v}_i^\top\mathbf{J}_{\mathcal{F}_{nn}}(\mathbf{A}_j; \mathbf{\theta}) - \mathbf{v}_i^\top \mathbf{J}_\mathcal{F}(\mathbf{A}_j)\|^2_2 
\Biggr]
\end{aligned}
\end{equation}

where $\mathbf{J}_\mathcal{F}(\mathbf{A})=\frac{\partial \mathcal{F}(\mathbf{A})}{\partial \mathbf{A}}$ and $\mathbf{J}_{\mathcal{F}_{nn}}(\mathbf{A})=\frac{\partial \mathcal{F}_{nn}(\mathbf{A})}{\partial \mathbf{A}}$ represent the Jacobians of the reservoir simulator and its neural network surrogate, respectively. The vectors $\{\mathbf{v}_i\}_{i=1}^{r}$ serve as the probing directions derived from the FIM, with $r$ specifying the number of rank. Finally, $\lambda$ is our regularization parameter. This objective function combines the $\ell_2$-norm misfit of the forward solutions between the reservoir simulator and the surrogate FNO, along with the $\ell_2$-norm misfit of the corresponding vector-Jacobian product. By jointly minimizing these terms over $N$ training samples, we drive the neural operator toward improved accuracy in both forward predictions and gradient estimation.

\subsubsection{Fourier Neural Operator \cite[]{li2020fourier}}

Among the various neural operators, such as the Graph Neural Operator (GNO) and Low-Rank Neural Operator (LNO), the Fourier Neural Operator (FNO) stands out for its computational efficiency \cite[]{li2020neural, li2020multipole}. By learning weights in the Fourier space, FNO transforms convolutional operations into simple multiplications, thereby significantly accelerating the solution of PDEs. This efficiency makes FNO a popular surrogate model for data-driven approaches to solving numerical PDEs \cite[]{lu2021learning, kovachki2023neural, karniadakis2021physics}. Especially, despite the initial training cost, FNO has been shown to generate approximate PDE solutions orders of magnitude faster than traditional numerical solvers \cite[]{li2020fourier}. That is, given an unseen spatial distribution of permeability $\mathbf{A}$, the trained FNO can near instantly produce time-dependent plume concentration snapshots via forward evaluation, i.e.,$\mathcal{F}_{nn}(\mathbf{A})$ \cite[]{li2020neural, wen2022u, zhang2022fourier}. Moreover, automatic differentiation provides access to gradients with respect to the FNO's input, further supporting downstream optimization tasks.

\begin{figure}
    \centering
    \includegraphics[width=0.49\linewidth]{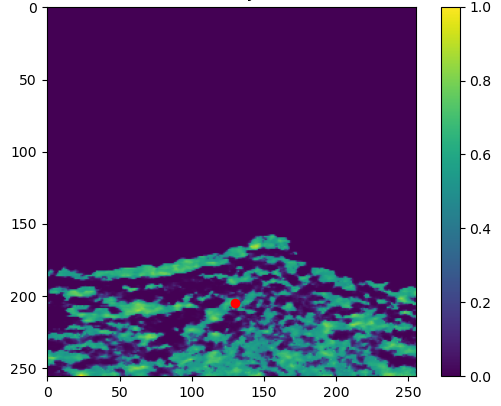}
        \includegraphics[width=0.48\linewidth]{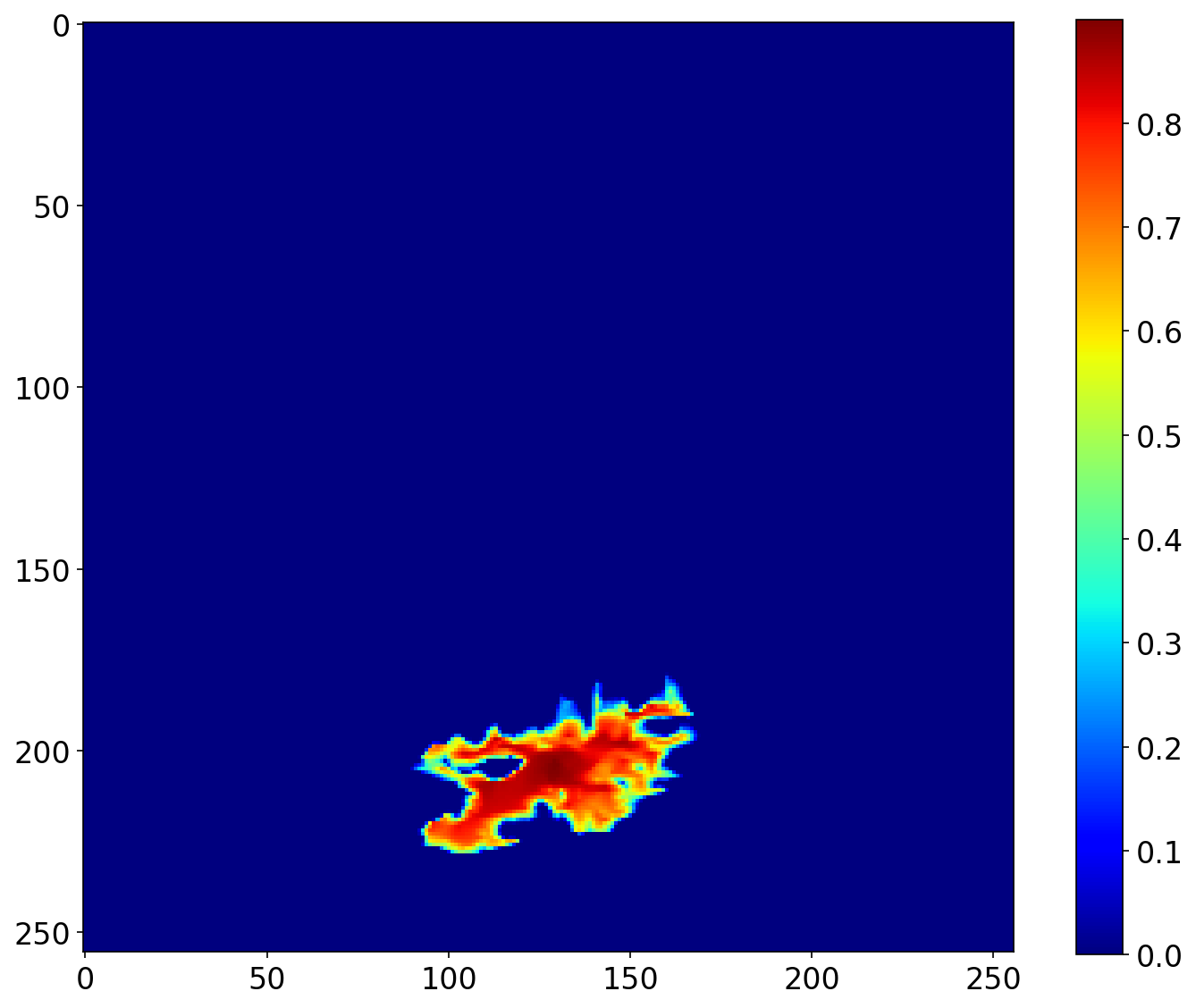}
    \caption{(Left) Example of a heterogeneous permeability model, $\mathbf{A}$, with injection location shown as red dot. Adapted from \cite[]{gahlot2024uncertainty} (Right) CO$_2$ plume concentration after 1,825 days of injection.}
    \label{fig:perm}
\end{figure}

\begin{figure}
    \centering
        \centering
    \includegraphics[width=1\linewidth]{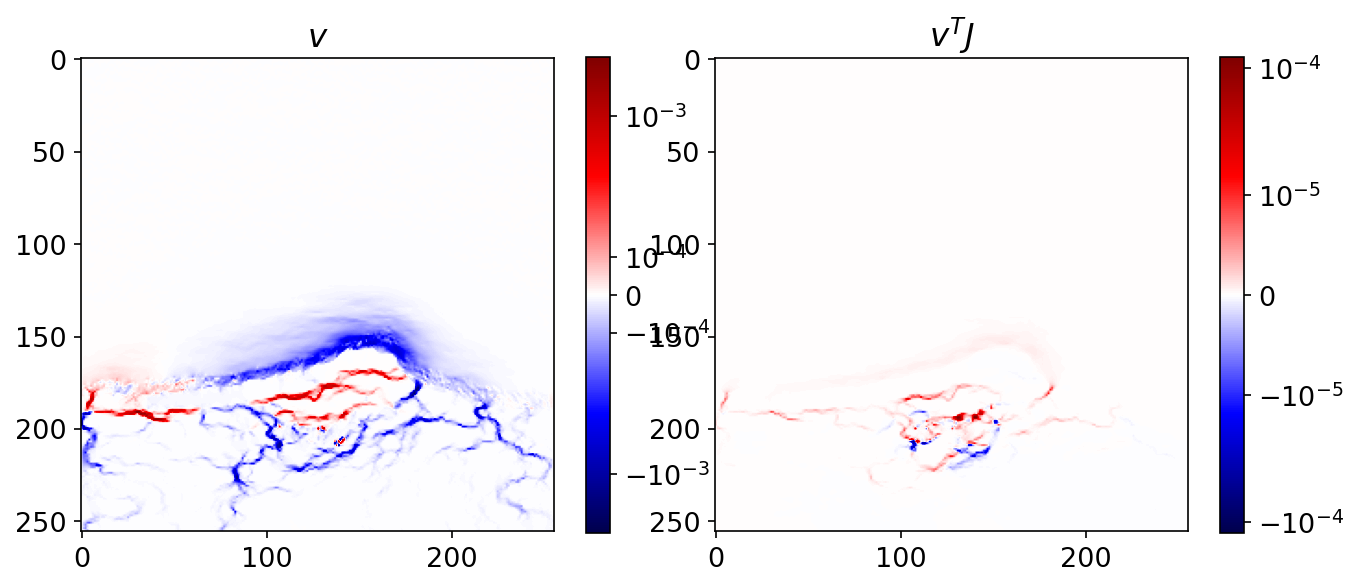}
    \caption{(Left) First eigenvector $\mathbf{v}$ of the Fisher Information Matrix (FIM) for a sample permeability model after 1,095 days of CO$_2$ plume evolution. (Right) Corresponding ground-truth vector–Jacobian product, $\mathbf{v}^\top\mathbf{J}$, computed using the reservoir simulator $\mathcal{F}$.}
    \label{fig:sample_eig}
\end{figure}

\subsubsection{Creating joint samples}

In the offline phase, we generate joint samples $\{\mathbf{A}_j, \mathbf{U}_j, \mathbf{v}_j, \mathbf{v}_j^\top\mathbf{J}\}_{j=1}^N$ to train our surrogate model. Here, $\mathbf{U} = \mathcal{F}(\mathbf{A})$ with $\mathbf{A}, \mathbf{U} \in \mathbb{R}^{d \times d}$, where $d$ denotes the number of grid points along each spatial dimension. The mapping $\mathcal{F}$ denotes the fluid-flow model implemented in \href{https://github.com/sintefmath/JutulDarcy.jl}{JutulDarcy.jl} \cite[]{jutuldarcy_ecmor_2024} and \href{https://github.com/slimgroup/JutulDarcyRules.jl}{JutulDarcyRules.jl} \cite[]{jutuldarcyrules}. For each sample, $\mathbf{v}_i$ is the $i$th eigenvector of the FIM, and $\mathbf{v}_i^\top \mathbf{J}$ is the corresponding vector–Jacobian product.

A major challenge in this process is the high computational cost of forming the FIM, which can sometimes be intractable. To address this, we consider an observation model $$
\mathbf{Y} = \mathcal{F}(\mathbf{A}) + \epsilon, \quad \epsilon \sim \mathcal{N}(0,\Sigma),
$$ which yields the following closed-form expression for the Fisher Information Matrix (FIM):
\begin{equation} \label{eq:pro_FIM} I(\mathbf{A}) = \left(\frac{\partial \mathcal{F}(\mathbf{A})}{\partial \mathbf{A}}\right)^T \Sigma^{-1} \left(\frac{\partial \mathcal{F}(\mathbf{A})}{\partial \mathbf{A}}\right), \end{equation} where $\mathcal{F}: \mathbb{R}^{d \times d} \to \mathbb{R}^{d \times d}$ is a deterministic mapping from the input permeability field to the CO$_2$ concentration.

To compute the first $r$ eigenvectors of $I(\mathbf{A})$, we extract the left singular vectors of $\mathbf{J}$, noting that the eigenvectors of $(\mathbf{v}^\top \mathbf{J})(\mathbf{v}^\top \mathbf{J})^\top$ coincide with the left singular vectors of $\mathbf{v}^\top \mathbf{J}$. This relies on the assumption that, in high-dimensional systems, the gradient information lies intrinsically in a low-dimensional subspace—making the Gauss–Newton Hessian $\mathbf{J}^\top \mathbf{J}$ amenable to low-rank approximation \cite[]{o2024derivative}.

Examples of $\mathbf{v}$ and the corresponding vector–Jacobian product $\mathbf{v}^\top \mathbf{J}$, computed using the proposed approach, are illustrated in Figure~\ref{fig:sample_eig}. The plot shows the first eigenvector $\mathbf{v}$ of the FIM and the corresponding Jacobian action $\mathbf{v}^\top \mathbf{J}$, based on the permeability model and CO$_2$ plume shown in Figure~\ref{fig:perm}. While $\mathbf{v}$ reveals the sensitivity structure with respect to the permeability field, $\mathbf{v}^\top \mathbf{J}$ highlights regions where the saturation has evolved, concentrating around the plume front. This approach avoids explicitly forming the full Gauss–Newton Hessian $\mathbf{J}^\top \mathbf{J}$, which would require $O(d^4)$ operations, and instead computes only $r$ vector–Jacobian products, reducing both time and memory complexity to $O(r \times d^2)$.

\subsection{NUMERICAL EXPERIMENTS}\label{results}

In our numerical experiments, we validate the performance of our proposed framework within a multiphase flow scenario by predicting the evolution of plume concentration from a given permeability field, represented through time-lapse snapshots. We assess both the output prediction accuracy and the fidelity of the directional derivative, $\mathbf{v}^\top \mathbf{J}$, as computed by the neural operator.

\subsubsection{Training setup}

In this work, we focus on evaluating the efficacy of our framework with a limited training dataset, consisting of 64 pairs of heterogenous permeability fields (shown in Figure~\ref{fig:perm}) and the corresponding time evolution of CO$_2$ concentration (example snapshot shown in Figure~\ref{fig:perm}). The permeability models are defined on a 256 $\times$ 256 grid with a grid spacing of 15 m in both the vertical and horizontal directions, and the associated porosity models are both derived from the work of \cite[]{gahlot2024uncertainty}. For the fluid-flow simulations, an injection well is placed on the middle part of the model, injecting supercritical CO$_2$ into the reservoir with the injection rate of $7 \times 10^{-3} \text{m}^3\text{/s}$. The time evolution of CO$_2$ concentration is modeled using \href{https://github.com/sintefmath/JutulDarcy.jl}{JutulDarcy.jl} \cite[]{jutuldarcy_ecmor_2024} and \href{https://github.com/slimgroup/JutulDarcyRules.jl}{JutulDarcyRules.jl} \cite[]{jutuldarcyrules} over a period of 5 years, with a time step of 1 year. This simulation produces five distinct CO$_2$ plume snapshots at different time steps for each permeability model. The Fourier Neural Operator (FNO) implementation employed in this study is based on the \href{https://developer.nvidia.com/physicsnemo}{NVIDIA Modulus framework}.

\subsubsection{Results and evaluation metrics}

\begin{table}[ht]
\centering
\begin{tabular}{>{\raggedright\arraybackslash}p{3cm}cc}
\hline
 & \textbf{FNO} & \textbf{DeFINO} \\
\hline
$\ell_2$-norm Misfit of Forward & $2.1440 \times 10^{-3}$ & $\mathbf{2.0725 \times 10^{-3}}$ \\
Relative $\ell_2$-norm Misfit of Gradient & $1.6426 \times 10^{-5}$ & $\mathbf{1.1739 \times 10^{-6}}$ \\
Reduced Gauss-Newton Error & $0.8554$ & $\mathbf{0.7697}$ \\
\hline
\end{tabular}
\caption{Comparison of FNO and DeFINO in gradient accuracy. Relative $\ell_2$-norm Misfit of Gradient is defined in Equation \ref{eq:L2}. Reduced Gauss–Newton Error (Equation \ref{eq:GNH}) measures the relative discrepancy (in a reduced subspace spanned by $V_r$) between the true and predicted Gauss–Newton Hessians (Adapted from \cite[]{o2024derivative}). Bold values indicate lower (i.e. better) errors.}
\label{tab:comparison}
\end{table}

After training both models under the same hyperparameter setting, we evaluate the effectiveness of the proposed DeFINO framework by assessing both forward prediction accuracy and gradient accuracy on a set of unseen test samples. Each test sample consists of a permeability model and five corresponding snapshots of CO$_2$ concentration obtained at intervals of 365 days using reservoir simulator. Following the procedure outlined in the Methodology section, form a test set comprising $\{\mathbf{A}_k, \mathbf{U}_k, \mathbf{v}_k, \mathbf{v}^\top_k\mathbf{J}\}_{k=1}^{100}$. 

\begin{figure}
    \centering
    \includegraphics[width=0.305\linewidth]{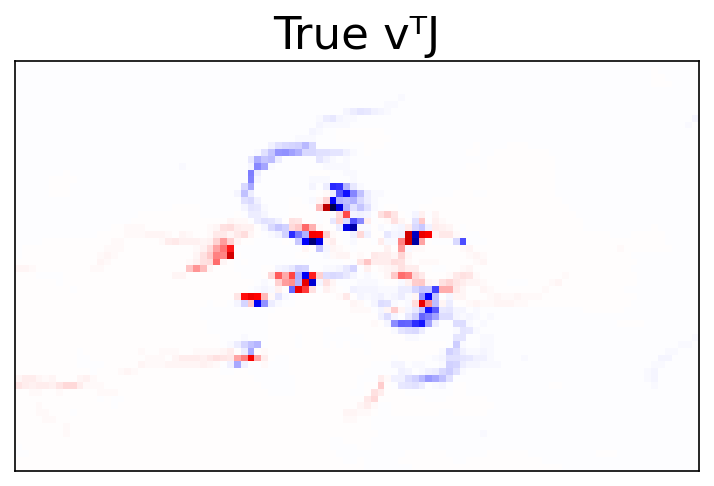}
    \includegraphics[width=0.30\linewidth]{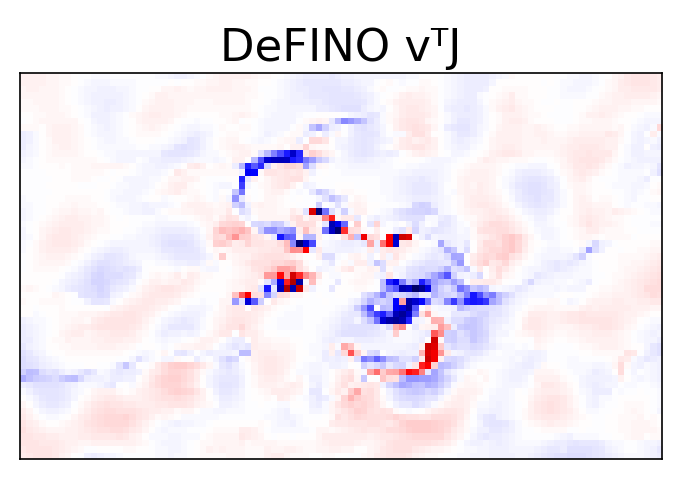}
    \includegraphics[width=0.37\linewidth]{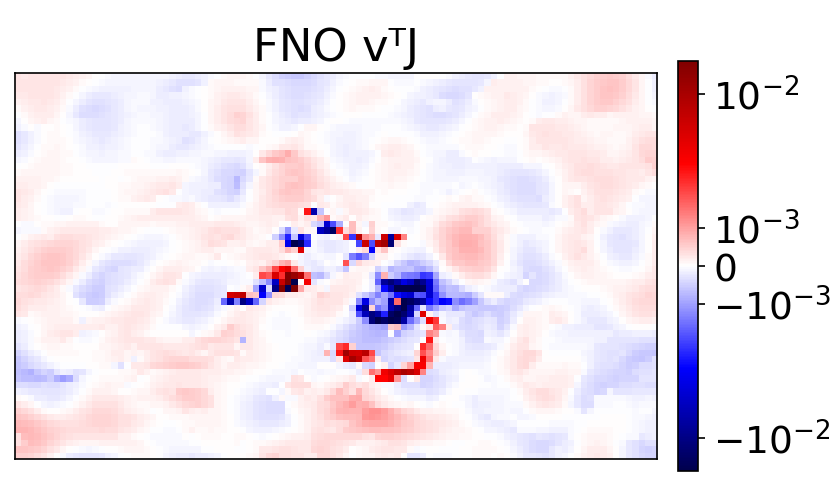}
    \caption{Close-up view of the gradient \(\mathbf{v}^\top\mathbf{J}\) after 365 days of CO\(_2\) injection. The first column shows the ground-truth gradient, the second column shows the gradient predicted by DeFINO, and the third column shows the gradient predicted by the baseline FNO. The relative \(\ell_2\)-norm misfit of DeFINO's \(\mathbf{v}^\top\mathbf{J}\) shown in the middle column is $0.0446$, compared to $0.1470$ for FNO shown in the right column.}
    \label{fig:eig_pred}
\end{figure}

We first compare the forward inference performance of a baseline FNO, trained solely with mean squared error (MSE) loss, against our proposed DeFINO model. For CO$_2$ plume prediction, DeFINO achieves a slightly improved $\ell_2$-norm misfit of $2.0725 \times 10^{-3}$ compared to the baseline FNO’s misfit of $2.1440 \times 10^{-3}$. This improvement in forward prediction suggests that incorporating gradient information during training enhances not only the model's ability to learn gradients but also its accuracy in capturing fluid plume evolution.

Beyond forward predictions, we also evaluate gradient accuracy, which is critical in inversion problems and uncertainty quantification tasks \cite[]{yin2023solving, o2022derivative}. Specifically, we assess the accuracy of the surrogate-derived gradient, $\mathbf{v}^\top\mathbf{J}_{\mathcal{F}_{nn}}$, compared to the true gradient computed via the reservoir simulator, $\mathbf{v}^\top\mathbf{J}_{\mathcal{F}}$, across the test samples. We quantify this accuracy using two metrics: the relative $\ell_2$-norm misfit of the gradients (Equation \ref{eq:L2}) and the reduced Gauss–Newton error (Equation \ref{eq:GNH}), which measures how well the surrogate approximates the true Gauss–Newton Hessian ($\mathbf{J}^\top \mathbf{J}$) along dominant parameter directions identified by the FIM \cite[]{o2024derivative}. Gauss–Newton Hessians are particularly valuable in inversion and optimization settings, as they provide curvature information essential for robust parameter estimation \cite[]{o2022derivative}.

\begin{equation}
    \label{eq:L2}
    \frac{\|\mathbf{V}^\top\mathbf{J}_{\mathcal{F}} - \mathbf{V}^\top\mathbf{J}_{\mathcal{F}_{nn}}\|_F^2}{\|\mathbf{V}^\top\mathbf{J}_{\mathcal{F}}\|_F^2}
\end{equation}

\begin{equation}
\label{eq:GNH}
\sqrt{\mathbb{E}_{\mathbf{A}} \left[\frac{\|\mathbf{V}_r^\top\bigl( \mathbf{J}_{\mathcal{F}}^\top  \mathbf{J}_{\mathcal{F}} -   \mathbf{J}_{\mathcal{F}_{nn}}^\top\mathbf{J}_{\mathcal{F}_{nn}}\bigr)\mathbf{V}_r\|_F^2}{\|\mathbf{V}_r^\top  \mathbf{J}_{\mathcal{F}}^\top\mathbf{J}_{\mathcal{F}}\mathbf{V}_r\|_F^2}\right]}
\end{equation}

Our results, summarized in Table~\ref{tab:comparison}, demonstrate that DeFINO significantly improves gradient estimation accuracy, achieving a relative $\ell_2$-norm misfit (Equation~\ref{eq:L2}) of $1.1739 \times 10^{-6}$ compared to the baseline FNO’s misfit of $1.6426 \times 10^{-5}$, an improvement of an order of magnitude. Furthermore, DeFINO yields a lower reduced Gauss–Newton error (Equation~\ref{eq:GNH}) of $0.7697$, compared to $0.8554$ for the FNO, highlighting its enhanced ability to approximate critical second-order information. As illustrated in Figure~\ref{fig:eig_pred}, the gradient predicted by DeFINO more closely resembles the structural features of the ground-truth gradient, capturing both upper and lower spherical regions, whereas the FNO prediction reflects only the lower part. Collectively, these results indicate that incorporating derivative-informed training guided by the Fisher Information Matrix substantially improves gradient fidelity without compromising forward prediction accuracy, making DeFINO particularly well-suited for inversion and other gradient-sensitive downstream applications.

\subsection{DISCUSSION AND CONCLUSION}\label{conclusion-and-discussion}

In this work, we introduced DeFINO, a reduced-order derivative-informed Neural Operator framework designed to enhance surrogate modeling for time-dependent multiphase flow problems. By incorporating derivative-based training via projections onto dominant eigen-directions of the FIM, DeFINO achieves improved accuracy in capturing gradients and Gauss–Newton Hessians, as demonstrated by our results in Table \ref{tab:comparison}. These improvements are particularly notable in scenarios with limited training data, reflecting realistic constraints often encountered in practice. Additionally, our approach reduces the computational cost of derivative information from $O(d^4)$ to $O(r \times d^2)$ by avoiding the explicit formation of full Jacobians. The enhanced derivative estimation and more accurate Hessian approximations provided by DeFINO make it a promising approach for Bayesian inversion, optimization under uncertainty, and robust forecasting of subsurface fluid-flow dynamics.

While these initial results are encouraging, further developments remain necessary for broader applicability and robustness. Future work will explore using the trained DeFINO model within a Bayesian inversion framework, enabling efficient estimation of reservoir permeability fields directly from observational data. Such inverse modeling can enhance uncertainty quantification, helping stakeholders better manage risks associated with CO$_2$ plume migration. Finally, assessing the generalization capability of DeFINO across diverse subsurface scenarios and extending the framework toward large-scale, continuous 3D monitoring workflows represent promising avenues for future research.

\subsection{ACKNOWLEDGEMENT}
During the preparation of the work, the authors used ChatGPT to refine sentence structures of the manuscript. After using this service, the authors reviewed and edited the content as needed and take full responsibiltiy for the content of the publication.

\bibliographystyle{IMAGE2025}  
\bibliography{abstract}

\begin{thebibliography}{}
\itemsep0pt

\bibitem[Benner et~al., 2015]{benner2015survey}
Benner, P., S. Gugercin, and K. Willcox,  2015, A survey of projection-based model reduction methods for parametric dynamical systems: SIAM review, {\bfseries 57}, 483--531.

\bibitem[Cai et~al., 2021]{cai2021physics}
Cai, S., Z. Mao, Z. Wang, M. Yin, and G.~E. Karniadakis,  2021, Physics-informed neural networks (pinns) for fluid mechanics: A review: Acta Mechanica Sinica, {\bfseries 37}, 1727--1738.

\bibitem[Chandra et~al., 2025]{chandra2025fourier}
Chandra, A., M. Koch, S. Pawar, A. Panda, K. Azizzadenesheli, J. Snippe, F.~O. Alpak, F. Hariri, C. Etienam, P. Devarakota, et~al.,  2025, Fourier neural operator based surrogates for $ co\_2 $ storage in realistic geologies: arXiv preprint arXiv:2503.11031.

\bibitem[Constantine, 2015]{constantine2015active}
Constantine, P.~G.,  2015, Active subspaces: Emerging ideas for dimension reduction in parameter studies: SIAM.

\bibitem[Constantine et~al., 2014]{constantine2014active}
Constantine, P.~G., E. Dow, and Q. Wang,  2014, Active subspace methods in theory and practice: applications to kriging surfaces: SIAM Journal on Scientific Computing, {\bfseries 36}, A1500--A1524.

\bibitem[Cui and Zahm, 2021]{cui2021data}
Cui, T., and O. Zahm,  2021, Data-free likelihood-informed dimension reduction of bayesian inverse problems: Inverse Problems, {\bfseries 37}, 045009.

\bibitem[Gahlot et~al., 2024]{gahlot2024uncertainty}
Gahlot, A.~P., R. Orozco, Z. Yin, and F.~J. Herrmann,  2024, An uncertainty-aware digital shadow for underground multimodal co2 storage monitoring: arXiv preprint arXiv:2410.01218.

\bibitem[Gao et~al., 2024]{gao2024adaptive}
Gao, Z., L. Yan, and T. Zhou,  2024, Adaptive operator learning for infinite-dimensional bayesian inverse problems: SIAM/ASA Journal on Uncertainty Quantification, {\bfseries 12}, 1389--1423.

\bibitem[Karniadakis et~al., 2021]{karniadakis2021physics}
Karniadakis, G.~E., I.~G. Kevrekidis, L. Lu, P. Perdikaris, S. Wang, and L. Yang,  2021, Physics-informed machine learning: Nature Reviews Physics, {\bfseries 3}, 422--440.

\bibitem[Kovachki et~al., 2023]{kovachki2023neural}
Kovachki, N., Z. Li, B. Liu, K. Azizzadenesheli, K. Bhattacharya, A. Stuart, and A. Anandkumar,  2023, Neural operator: Learning maps between function spaces with applications to pdes: Journal of Machine Learning Research, {\bfseries 24}, 1--97.

\bibitem[Li et~al., 2020a]{li2020fourier}
Li, Z., N. Kovachki, K. Azizzadenesheli, B. Liu, K. Bhattacharya, A. Stuart, and A. Anandkumar,  2020a, Fourier neural operator for parametric partial differential equations: arXiv preprint arXiv:2010.08895.

\bibitem[Li et~al., 2020b]{li2020neural}
--------, 2020b, Neural operator: Graph kernel network for partial differential equations: arXiv preprint arXiv:2003.03485.

\bibitem[Li et~al., 2020c]{li2020multipole}
Li, Z., N. Kovachki, K. Azizzadenesheli, B. Liu, A. Stuart, K. Bhattacharya, and A. Anandkumar,  2020c, Multipole graph neural operator for parametric partial differential equations: Advances in Neural Information Processing Systems, {\bfseries 33}, 6755--6766.

\bibitem[Li et~al., 2024]{li2024physics}
Li, Z., H. Zheng, N. Kovachki, D. Jin, H. Chen, B. Liu, K. Azizzadenesheli, and A. Anandkumar,  2024, Physics-informed neural operator for learning partial differential equations: ACM/JMS Journal of Data Science, {\bfseries 1}, 1--27.

\bibitem[Lu et~al., 2021]{lu2021learning}
Lu, L., P. Jin, G. Pang, Z. Zhang, and G.~E. Karniadakis,  2021, Learning nonlinear operators via deeponet based on the universal approximation theorem of operators: Nature machine intelligence, {\bfseries 3}, 218--229.

\bibitem[M{\o}yner, 2024]{jutuldarcy_ecmor_2024}
M{\o}yner, O.,  2024, Jutuldarcy.jl - a fully differentiable high-performance reservoir simulator based on automatic differentiation: {\bfseries 2024}, 1--9.

\bibitem[O'Leary-Roseberry et~al., 2024]{o2024derivative}
O'Leary-Roseberry, T., P. Chen, U. Villa, and O. Ghattas,  2024, Derivative-informed neural operator: an efficient framework for high-dimensional parametric derivative learning: Journal of Computational Physics, {\bfseries 496}, 112555.

\bibitem[O’Leary-Roseberry et~al., 2022]{o2022derivative}
O’Leary-Roseberry, T., U. Villa, P. Chen, and O. Ghattas,  2022, Derivative-informed projected neural networks for high-dimensional parametric maps governed by pdes: Computer Methods in Applied Mechanics and Engineering, {\bfseries 388}, 114199.

\bibitem[Park et~al., 2024]{park2024dynamical}
Park, J., N. Yang, and N. Chandramoorthy,  2024, When are dynamical systems learned from time series data statistically accurate?: arXiv preprint arXiv:2411.06311.

\bibitem[Qiu et~al., 2024]{qiu2024derivative}
Qiu, Y., N. Bridges, and P. Chen,  2024, Derivative-enhanced deep operator network: arXiv preprint arXiv:2402.19242.

\bibitem[Quarteroni et~al., 2015]{quarteroni2015reduced}
Quarteroni, A., A. Manzoni, and F. Negri,  2015, Reduced basis methods for partial differential equations: an introduction: Springer, {\bfseries 92}.

\bibitem[Rosofsky et~al., 2023]{rosofsky2023applications}
Rosofsky, S.~G., H. Al~Majed, and E. Huerta,  2023, Applications of physics informed neural operators: Machine Learning: Science and Technology, {\bfseries 4}, 025022.

\bibitem[Sheriffdeen et~al., 2019]{sheriffdeen2019accelerating}
Sheriffdeen, S., J.~C. Ragusa, J.~E. Morel, M.~L. Adams, and T. Bui-Thanh,  2019, Accelerating pde-constrained inverse solutions with deep learning and reduced order models: arXiv preprint arXiv:1912.08864.

\bibitem[Wen et~al., 2022]{wen2022u}
Wen, G., Z. Li, K. Azizzadenesheli, A. Anandkumar, and S.~M. Benson,  2022, U-fno—an enhanced fourier neural operator-based deep-learning model for multiphase flow: Advances in Water Resources, {\bfseries 163}, 104180.

\bibitem[Wu et~al., 2023]{wu2023large}
Wu, K., T. O’Leary-Roseberry, P. Chen, and O. Ghattas,  2023, Large-scale bayesian optimal experimental design with derivative-informed projected neural network: Journal of Scientific Computing, {\bfseries 95}, 30.

\bibitem[Wu et~al., 2024]{wu2024uncertainty}
Wu, T., W. Neiswanger, H. Zheng, S. Ermon, and J. Leskovec,  2024, Uncertainty quantification for forward and inverse problems of pdes via latent global evolution: Proceedings of the AAAI Conference on Artificial Intelligence, 320--328.

\bibitem[Yin et~al., 2024]{jutuldarcyrules}
Yin, Z., G. Bruer, and M. Louboutin,  2024, slimgroup/jutuldarcyrules.jl: v0.2.8.

\bibitem[Yin et~al., 2023]{yin2023solving}
Yin, Z., R. Orozco, M. Louboutin, and F.~J. Herrmann,  2023, Solving multiphysics-based inverse problems with learned surrogates and constraints: Advanced Modeling and Simulation in Engineering Sciences, {\bfseries 10}, 14.

\bibitem[Yin et~al., 2022]{yin2022learned}
Yin, Z., A. Siahkoohi, M. Louboutin, and F.~J. Herrmann,  2022, Learned coupled inversion for carbon sequestration monitoring and forecasting with fourier neural operators: Second International Meeting for Applied Geoscience \& Energy, Society of Exploration Geophysicists and American Association of Petroleum~…, 467--472.

\bibitem[Zahm et~al., 2022]{zahm2022certified}
Zahm, O., T. Cui, K. Law, A. Spantini, and Y. Marzouk,  2022, Certified dimension reduction in nonlinear bayesian inverse problems: Mathematics of Computation, {\bfseries 91}, 1789--1835.

\bibitem[Zhang et~al., 2022]{zhang2022fourier}
Zhang, K., Y. Zuo, H. Zhao, X. Ma, J. Gu, J. Wang, Y. Yang, C. Yao, and J. Yao,  2022, Fourier neural operator for solving subsurface oil/water two-phase flow partial differential equation: Spe Journal, {\bfseries 27}, 1815--1830.

\end{thebibliography}

\end{document}